\documentclass[hidelinks]{article}
\usepackage{lmodern}

\usepackage[T1]{fontenc}
\usepackage[utf8]{inputenc}
\usepackage[english]{babel}
\usepackage{amsmath}
\usepackage{amsthm}
\usepackage{amssymb}
\usepackage{graphicx}
\usepackage[round]{natbib}
\usepackage[] {hyperref}
\usepackage[] {xcolor}
\usepackage[realmainfile]{currfile}
\usepackage{geometry}
\title{How the interpolation of life tables affects the decomposition of life insurance surplus}
\author{Mintodê Nicodème Atchadé\footnote{National Higher School of Mathematical Engineering and Modeling, National University of Sciences, Technologies, Engineering and Mathematics, Abomey, Benin Republic} \and Marcus C. Christiansen\footnote{Institut für Mathematik, Carl von Ossietzky Universität, 26129, Oldenburg, Germany, E-mail: marcus.christiansen@uol.de
} \and Friedrich Hubalek\footnote{TU Wien, Financial and Actuarial Mathematics, Wiedner Hauptstraße 8 / 105-1, A-1040 Vienna, Austria, E-mail: fhubalek@fam.tuwien.ac.at
} \and Gero Junike\footnote{Corresponding author. Institut f\"ur Mathematik, Ludwig-Maximilians-Universit\"at Munich, Germany, E-mail: gero.junike@math.lmu.de, ORCID: 0000-0001-8686-2661}}
\date{\today}

\begin{document}
\title{How the interpolation of life tables affects the decomposition of life insurance surplus}
\maketitle
\begin{abstract}
The surplus of a life insurance policy depends on both systematic changes in mortality risk and financial changes. We propose to decompose the surplus by the axiomatically justified IASU decomposition,
which is a continuous time version of the Shapley value. However, life tables are not updated continuously, but rather, only once per year. In this yearly update cycle of the life tables, we apply different interpolation methods to perform the IASU decomposition and analyze the effects of these methods on the surplus decomposition. Our results show that Lee-Carter and linear interpolation yield almost identical
decompositions, whereas constant approximations results in substantially different decompositions. As
a consequence, reporting standards and regulators should clarify how to interpolate mortality risks.
\end{abstract}

\section{Introduction}

Life insurance involves not only biometric risks but also financial risks. Examples include traditional life insurance policies, which are subject to interest rate risk, as well as equity-linked insurance contracts. Due to various reporting standards and regulatory rules\footnote{See for instance, the market consistent embedded value (MCEV) reporting
principles of the CFO Forum (2016), the International Financial Reporting
Standard (IFRS) 17, issued by the International Accounting Standards
Board (IASB) (2017), the insurance regulation of the European Union
(EU) (2015) and the MindZV (Mindestzuführungsverordnung) issued by
the BaFin (Germany's financial supervisory authority).}, the surplus of these products needs to be decomposed into a financial
and a biometric part. Let us assume the surplus can be written as
$f(D_{1},S_{1})-f(D_{0},S_{0})$, where $D_{t}$ and $S_{t}$ are
risk factors describing financial and mortality risk, respectively,
at time $t\in\{0,1\}$, and $f:\mathbb{R}^{2}\to\mathbb{R}$ is a
suitable function. The industrial standard to solve this problem is
defined as follows, see e.g., \citet[Satz 11.6]{HM99} or \citet{candland2014profit},
\begin{equation}
\delta^{D}=f(D_{1},S_{0})-f(D_{0},S_{0}),\quad\delta^{S}=f(D_{1},S_{1})-f(D_{1},S_{0}).\label{eq:KontributionFormula}
\end{equation}
The terms $\delta^{D}$ and $\delta^{S}$ are interpreted as the contributions
of the financial and the mortality risk factors to the surplus, respectively.
The formula~(\ref{eq:KontributionFormula}) is called \emph{sequential
updating }(SU) or \emph{waterfall} decomposition\emph{. }It is easy
to see that $\delta^{D}+\delta^{S}$ equals the total surplus. However,
the formula (\ref{eq:KontributionFormula}) is somewhat arbitrary.
For example, why is $D$ updated first instead of $S$? This topic
has been discussed in detail in the context of insurance, for example,
in \citet{CJ22,christiansen2023decomposition,CJS}. Real-world examples
demonstrate that different update orders and frequencies lead to
different contributions, see \citet{flaig2024profit}. To circumvent
this issue, we follow \citet{CJS} and propose decomposing the surplus
by the infinitesimal average sequential updating (IASU) decomposition,
which is a time dynamic version of the well-known Shapley value: the
IASU decomposition can be approximated by an iterative Shapley value.
The IASU decomposition takes into account the whole information of
the risk factors between two reporting dates. Note that financial
information, e.g., bond or equity prices, is updated frequently, e.g.,
daily. Conversely, information on biometric risks, such as life tables,
is usually only updated yearly. There are some exceptions: for example, mortality data is also available in some countries on a monthly or even weekly basis, but relevant covariates such as age are only available in aggregated form. Nevertheless, there will still be a discrepancy between the update frequency of financial and biometric data. In this  paper, we exploratory examine yearly updates for mortality data. 

In order to apply the IASU decomposition,
one must first interpolate information on biometric risks for fractional
years.
This letter makes the following main contribution: For illustrative
purposes, we price a portfolio of life insurance contracts using German
life tables and government bond prices. Then, we decompose the surplus
into a biometric part and a financial part. Bond prices are updated
daily, but life tables are only updated yearly. To decompose the portfolio
using all available information, we interpolate the life tables. We
use linear interpolation and harmonic interpolation, as proposed by
\citet{lee1992modeling}, as well as constant approximation. Our empirical
analysis shows that these different interpolation techniques have
a significant impact on the contribution of the two risk factors:
biometric and financial risk.

Our findings conclude that the regulator needs to improve the regulation
by stating more precisely how to obtain the contributions of the financial
and biometric risk factors. We recommend the IASU decomposition, or
the iterative Shapley value as an approximation, since they have sound
theoretical justifications (see \citet{CJS}). Additionally, it must
be clarified how to estimate biometric risks from life tables for
fractional years.

This letter is structured as follows: In Section\ref{sec:Life-insurance-contract},
we price a simple life insurance contract. In Section\ref{sec:Decomposition},
introduces the IASU decomposition and explains how to use it in practice.
In Section~\ref{sec:Numerical-experiments}, we price a life insurance
contract using German life tables and German government Bonds prices.
We empirically demonstrate that different interpolation techniques
for the life tables result in significantly different outcomes. 

\section{\protect\label{sec:Life-insurance-contract}Life insurance contract}

Let $(\Omega,\mathcal{F},(\mathcal{F}_{t})_{t\geq0},\mathbb{P})$
be a filtered probability space satisfying the usual conditions. Let
$T\in\mathbb{N}$. The measure $\mathbb{P}$ denotes the physical
measure. 

We consider an individual life insurance contract with maturity $T>0$, issued at time $0$ with a starting age of $x \in \mathbb{N}$ of the insured. Time $0$ corresponds to the last day of a specific year,
e.g., 31/12/2004. The insurance company pays a fixed amount of $N$
EUR to the insured if the insured is still alive at time $T$ and
nothing otherwise. We assume that the individual contract is part of a large life insurance portfolio, so that the unsystematic mortality risk is well diversified. We try to estimate a (minimum) fair premium
for this life insurance.

By $\tau(x)$ we denote the remaining life time of a person aged $x$
in calendar year zero. Let 
\begin{equation}
S_{t}:=\mathbb{E}\left[1_{\{\tau(x)>T\}}|\mathcal{F}_{t}\right],\quad t\in[0,T]\label{eq:A}
\end{equation}
be the conditional probability that an insured survives the contract,
given the demographic information at time $t$. 

Let $D_t$ be the discounting factor at time $t$ for a payment at time $T$. Then the product $\pi_t =D_t S_t$ gives the value of the life insurance contract at time $t$. 

We assume that $D_{t}$ is given by a financial market. Next, we estimate
$S_{t}$: Let
\[
p_{t}(k,k+1)=\frac{\mathbb{E}\left[1_{\{\tau(x)>k+1\}}|\mathcal{F}_{t}\right]}{\mathbb{E}\left[1_{\{\tau(x)>k\}}|\mathcal{F}_{t}\right]},\quad t\in[0,T],\quad k\in\mathbb{N}_{0},
\]
which is the conditional probability that $\tau(x)>k+1$ given that
$\tau(x)>k$ and given the demographic information at time $t$. We
assume that all insured are alive at the starting time of the contract,
i.e., $\mathbb{P}\big(\tau(x)>0\big)=1$, which implies that 
\[
S_{t}=p_{t}(0,1)p_{t}(1,2)\cdot\cdot\cdot p_{t}(T-1,T),\quad t\in[0,T],\quad\mathbb{P}-a.s.
\]
From a life table, we let $q_{x+k}^{\ell}$ be the one-year death
probability of a person aged $x+k$ as observed in calendar year $\ell$.
For example, when year zero corresponds to the last day of the calendar
year 2004, $q_{45}^{1}$, is the one-year death probability in 30/12/2005
of a person aged 45 in 30/12/2005. We estimate $p_{t}(k,k+1)$, $t\in[0,T]$,
$k\in\mathbb{N}_{0}$, by 
\[
\begin{cases}
1-q_{x+k}^{k+1} & ,k+1\leq t\\
1-q_{x+k}^{t} & ,t<k+1.
\end{cases}
\]
That is, if the corresponding biometric information is available,
i.e., whenever $k+1\leq t$, we use the life table of the age of interest,
i.e., $1-q_{x+k}^{k+1}$. Otherwise, we use the last observed life
table, i.e., $1-q_{x+k}^{t}$.

However, for $t\notin\mathbb{N}_{0}$ the term $1-q_{x+k}^{t}$ is
not available because life tables are updated on a yearly basis and
therefore we use interpolation: 
\[
q_{x+k}^{t}=\text{interpolation between }q_{x+k}^{\lfloor t\rfloor}\text{ and }q_{x+k}^{\lceil t\rceil}\text{ for }t\notin\mathbb{N}_{0}.
\]
 We study the following interpolation method, which is proposed by
\citet{lee1992modeling}
\[
q_{x+k}^{t}=\left(q_{x+k}^{\lfloor t\rfloor}\right)^{\lceil t\rceil-t}\left(q_{x+k}^{\lceil t\rceil}\right)^{t-\lfloor t\rfloor},\quad t\notin\mathbb{N}_{0},\quad0<t<k+1\leq T.
\]
We also propose to use linear interpolation: 
\[
q_{x+k}^{t}=(\lceil t\rceil-t)q_{x+k}^{\lfloor t\rfloor}+(t-\lfloor t\rfloor)q_{x+k}^{\lceil t\rceil},\quad t\notin\mathbb{N}_{0},\quad0<t<k+1\leq T,
\]
and we consider constant approximation, i.e., 
\[
q_{x+k}^{t}=q_{x+k}^{\lfloor t\rfloor},\quad t\notin\mathbb{N}_{0},\quad0<t<k+1\leq T.
\]
Interpolation methods for fractional ages are discussed in \citet[Sec. 3.6]{newton1997actuarial}
and \citet[Sec. 2.6]{gerber2013life}.

\section{\protect\label{sec:Decomposition}Decomposition}

As explained in Section~\ref{sec:Life-insurance-contract}, the value $\pi_t$ of the life insurance contract at time $t$ is given by the product
$\pi_t = S_t D_t$. of the discount factor $D_t$ and the survival probability $S_t$.

Due to movements in the yield curves and demographic changes, the
price typically changes over every business day generating a P\&L (profit and loss).
Due to regulatory requirements, one is interested in the decomposition
of the surplus into a financial part and a biometric part. Assuming
that $(D_{t})_{t\geq0}$ and $(S_{t})_{t\geq0}$ are semimartingales,
we decompose the surplus in $[0,t]$ of the life insurance portfolio,
using Itô's lemma, by
\begin{align}
\pi_{t}-\pi_{0} & =D_{t}S_{t}-D_{0}S_{0}\nonumber \\
 & =\left\{ \int_{0}^{t}S_{u-}dD_{u}+\frac{1}{2}[S,D]_{t}\right\} +\left\{ \int_{0}^{t}D_{u-}dS_{u}+\frac{1}{2}[S,D]_{t}\right\} ,\quad t\in[0,T].\label{eq:Ito}
\end{align}
The first term at the right-hand side of Equation~(\ref{eq:Ito})
is interpreted as the contribution of the surplus by changes of the
discount factor (due to movements in the financial market) and the
second term is interpreted as the contribution of the surplus by shifts
in the survival probabilities (due to demographic changes). Equation~(\ref{eq:Ito})
is the so called \emph{infinitesimal average sequential updating}
(IASU) decomposition. It is the only decomposition taking the whole
available information into account and satisfying many desirable properties,
see \citet{CJS}. The IASU decomposition
can be approximated in practical applications by the \emph{average
sequential updating} (ASU) decomposition, which is also known as the
\emph{Shapley value}. The ASU decomposition is very easy to implement,
for a straightforward introduction see \citet{flaig2024profit}. Next,
we explain the ASU decomposition in some details: we divide the interval
$[0,t]$ into $L(t)\in\mathbb{N}$ subintervals (e.g. daily time steps),
i.e., we let 
\[
\{0=\sigma_{0}<\sigma_{1}<...<\sigma_{L(t)}=t\}
\]
and define

\[
\delta_{t}^{D}=\frac{1}{2}\sum_{\ell=0}^{L(t)}\left(S_{\sigma_{\ell}}D_{\sigma_{\ell+1}}-S_{\sigma_{\ell}}D_{\sigma_{\ell}}\right)+\frac{1}{2}\sum_{\ell=0}^{L(t)}\left(S_{\sigma_{\ell+1}}D_{\sigma_{\ell+1}}-S_{\sigma_{\ell+1}}D_{\sigma_{\ell}}\right)
\]
and
\[
\delta_{t}^{S}=\frac{1}{2}\sum_{\ell=0}^{L(t)}\left(S_{\sigma_{\ell+1}}D_{\sigma_{\ell+1}}-S_{\sigma_{\ell}}D_{\sigma_{\ell+1}}\right)+\frac{1}{2}\sum_{\ell=0}^{L(t)}\left(S_{\sigma_{\ell+1}}D_{\sigma_{\ell}}-S_{\sigma_{\ell}}D_{\sigma_{\ell}}\right).
\]
The term $\delta_{t}^{D}$ is interpreted as the contribution of changes
in the discount factor $D$ to the surplus. To obtain $\delta_{t}^{D}$,
we freeze $S$ in each subinterval $(\sigma_{\ell},\sigma_{\ell+1}]$
at $\sigma_{\ell}$ and at $\sigma_{\ell+1}$ and only allow $D$
to vary between $\sigma_{\ell}$ and $\sigma_{\ell+1}$. Then we take
the average, which explains the factor $\frac{1}{2}$. Then we sum
over all subintervals. Similarly, $\delta_{t}^{S}$ is interpreted
as contribution of shifts in the survival probabilities to the surplus.
Applying telescoping series, one can see $\pi_{t}-\pi_{0}=\delta_{t}^{D}+\delta_{t}^{S}$
for any $L(t)\in\mathbb{N}$. The processes $\delta_{t}^{D}$ and
$\delta_{t}^{S}$ converge in probability to the corresponding terms
in Equation~(\ref{eq:Ito}), see \citet[Thm. 4.7]{CJS}.

\section{\protect\label{sec:Numerical-experiments}Numerical experiments}

\begin{figure}
\begin{centering}
\begin{tabular}{cc}
\includegraphics[height=6.5cm]{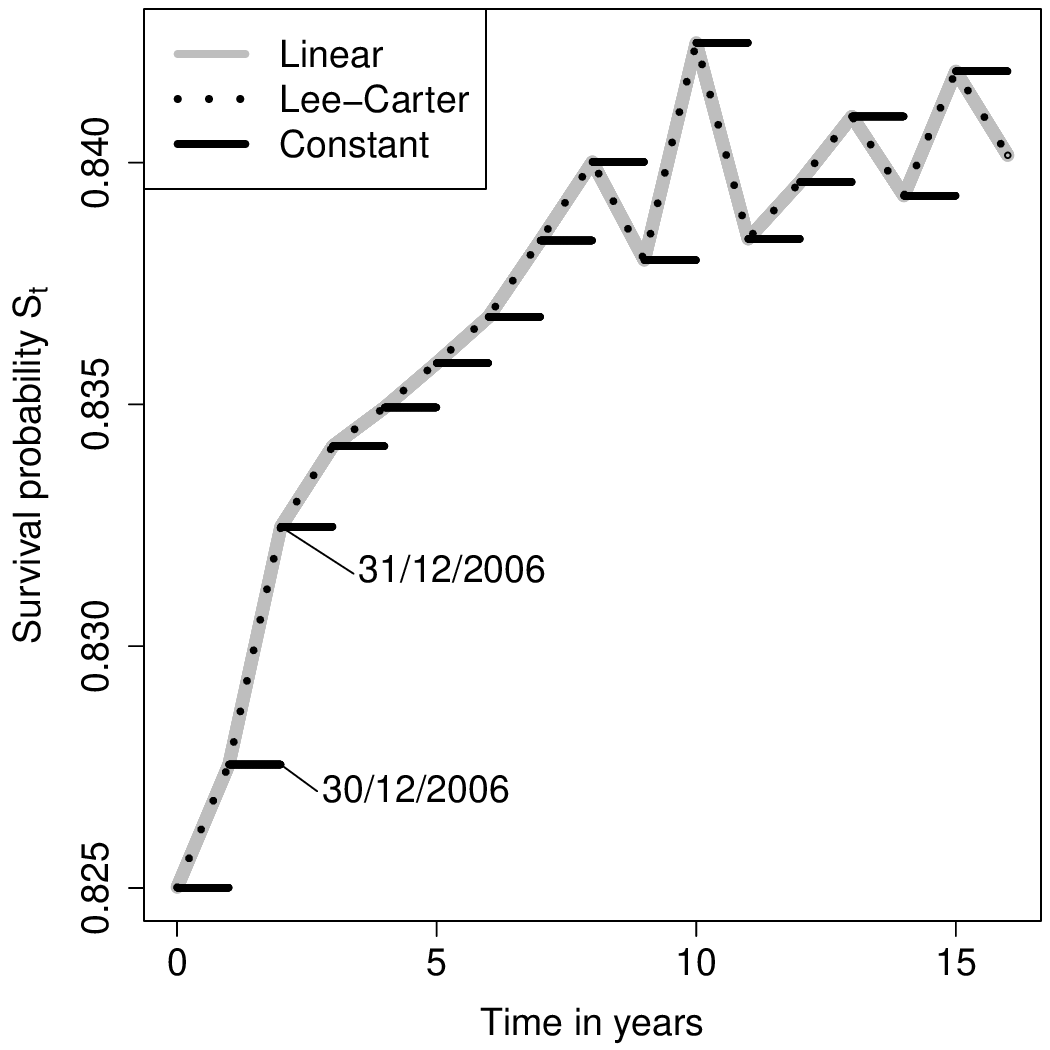} & \includegraphics[height=6.5cm]{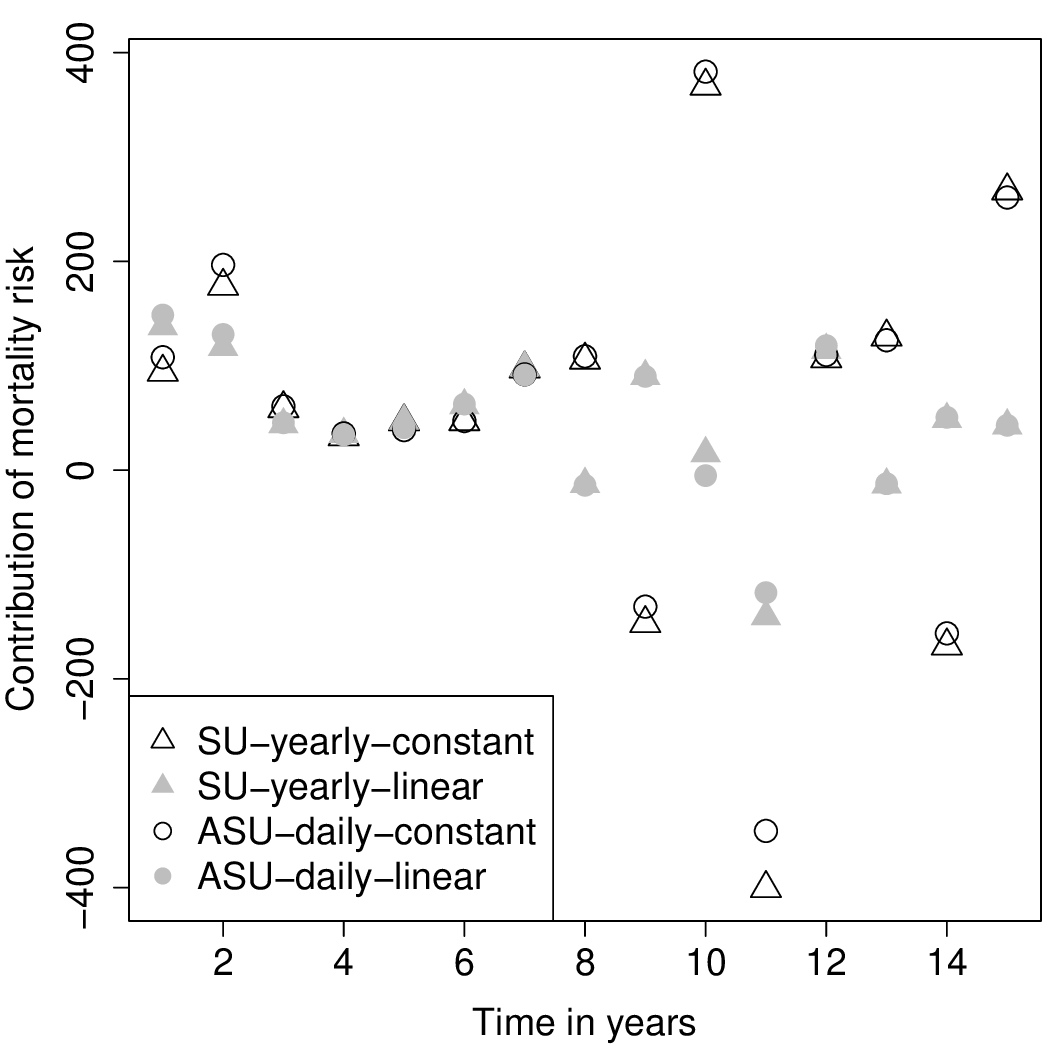}\tabularnewline
Panel A & Panel B\tabularnewline
\end{tabular}
\par\end{centering}
\caption{\protect\label{fig:1}Survival probability $S_{t}$ and contribution
of mortality risk estimated from life tables from 2005 - 2020 using
different interpolation methods for fractional years. Panel A: Survival
probability $S_{t}$. Panel B: contribution of mortality risk for
$N=100\,000$ using yearly SU or ASU with daily data. The contribution
is calculated every year and the reporting date is set to the middle of the year.}
\end{figure}

Exemplary, we consider a life insurance contract starting
in beginning of January 2005 with a maturity of 25 years, sold to
a person aged 45 years at the beginning of the contract. We price the
contract every day during 2005 - 2020. That is, $x=45$, $t\in[0,16)$
and $T=25$. To do so, we need to estimate the discount factor $D_{t}$
and the survival probability $S_{t}$, see Equation~(\ref{eq:A}).
In order to estimate $p_{t}(k,k+1)$, we use German life tables (both
sex) obtained from \url{mortality.org}. We construct the discount
factor $D_{t}$ using yields curves of German government bonds obtained
from the European central bank (\url{data.ecb.europa.eu/data/datasets/YC}).
For daily data, the yield curves with maturities 30y, 20y, 10y and
5y are available. We use linear interpolation to obtain the yield
for maturities between these years. In particular,
\[
D_{t}=\left(1+\frac{r(t,T-t)}{100}\right)^{-(T-t)},
\]
where $r(t,T-t)$ is the interest rate earned from a German zero coupon
Government bond issued at time $t$ with maturity at time $T$. We
apply Lee-Carter, linear interpolation and constant approximation
in order to estimate $S_{t}$ for daily data. Panel A in Figure \ref{fig:1}
shows the survival probabilities $S_{t}$ for fractional years. There
is very little difference between the Lee-Carter method and linear
interpolation. As expected, the constant approximation coincides with
the Lee-Carter and linear interpolations only at the end of each year.

As described in Section \ref{sec:Decomposition}, we compute the ASU
decomposition using daily data, i.e., we set $L(t)$ to the number
of business days in $[0,t]$. We also apply the SU decomposition,
as defined in Equation (\ref{eq:KontributionFormula}), using only
yearly data. 

The different decomposition principles and interpolation methods produce similar results, if the reporting date is the last day
of the year. In this case, the SU decomposition is independent of
the choice of interpolation method because it uses yearly data. In
this particular example, we also observe that the ASU decomposition
with daily data and the SU decomposition with yearly data do not differ
significantly. However, this is rather an exception than the rule,
compare with \citet{flaig2024profit}.

However, if the reporting date is different, the difference between the constant
approximation and the Lee-Carter/linear interpolation can be significant.
The worst case occurs when the penultimate day of the year is chosen
as the reporting date (e.g., 30/12/2006), which is the day before
jump of the constant approximation. Panel B in Figure \ref{fig:1}
shows the contribution of the mortality risk for all years using the
ASU and the SU decompositions with linear interpolation and constant
approximation, with the reporting date set to the middle of each year.
For example, the first reporting period considers the time frame from 4/7/2005
to 3/7/2006. We still observe that the difference between ASU and
SU is small. However, for some years - e.g., years nine, ten and eleven
of the contract - the difference between the constant approximation
and the linear interpolation is significant. For instance, in year
ten, the contribution of the mortality risk using the linear interpolation
is nearly zero, whereas the contribution using the constant approximation
is almost 400 euros per insured person. Consequently, the reporting standards and the regulators should
define more precisely how to decompose contracts involving financial
and biometric risks, as well as which interpolation methods to use
for infrequently observable time series, such as life tables.

\section*{Declarations}
\textbf{Conflict of interest:} The authors report no Conflict of interest
to declare.\\
\textbf{Data availability:} All of the data used in this research
is publicly available. URLs are provided in the letter.\\
\textbf{Funding Declaration:} This research was partially funded by the AIMS-DFG (African Institutes for Mathematical Sciences and the Deutsche Forschungsgemeinschaft) Collaboration Visits Programme in the Mathematical Sciences.

\bibliographystyle{plainnat}
\bgroup\inputencoding{utf8}\bibliography{biblio_surlife1}
\egroup

\end{document}